\def \SAIT #1 #2 {{\em Mem.\ Soc.\ Astron.\ It.\/} {\bf #1}, #2}
\def \MESS #1 #2 {{\em The Messenger\/} {\bf #1}, #2}
\def \ASTRNACH #1 #2 {{\em Astron. Nach.\/} {\bf #1}, #2}
\def \AAP #1 #2 {{\em Astron. Astrophys.\/} {\bf #1}, #2}
\def \AAL #1 #2 {{\em Astron. Astrophys. Lett.\/} {\bf #1}, L#2}
\def \AAR #1 #2 {{\em Astron. Astrophys. Rev.\/} {\bf #1}, #2}
\def \AAS #1 #2 {{\em Astron. Astrophys. Suppl. Ser.\/} {\bf #1}, #2}
\def \AJ #1 #2 {{\em Astron. J.\/} {\bf #1}, #2}
\def \ANNREV #1 #2 {{\em Ann. Rev. Astron. Astrophys.\/} {\bf #1}, #2}
\def \APJ #1 #2 {{\em Astrophys. J.\/} {\bf #1}, #2}
\def \APJL #1 #2 {{\em Astrophys. J. Lett.\/} {\bf #1}, L#2}
\def \APJS #1 #2 {{\em Astrophys. J. Suppl.\/} {\bf #1}, #2}
\def \APSS #1 #2 {{\em Astrophys. Space Sci.\/} {\bf #1}, #2}
\def \ASR #1 #2 {{\em Adv. Space Res.\/} {\bf #1}, #2}
\def \BAIC #1 #2 {{\em Bull. Astron. Inst. Czechosl.\/} {\bf #1}, #2}
\def \JSQRT #1 #2 {{\em J. Quant. Spectrosc. Radiat. Transfer\/} {\bf #1}, #2}
\def \MN #1 #2 {{\em Mon. Not. R. Astr. Soc.\/} {\bf #1}, #2}
\def \MEM #1 #2 {{\em Mem. R. Astr. Soc.\/} {\bf #1}, #2}
\def \PLR #1 #2 {{\em Phys. Lett. Rev.\/} {\bf #1}, #2}
\def \PASJ #1 #2 {{\em Publ. Astron. Soc. Japan\/} {\bf #1}, #2}
\def \PASP #1 #2 {{\em Publ. Astr. Soc. Pacific\/} {\bf #1}, #2}
\def \NAT #1 #2 {{\em Nature\/} {\bf #1}, #2}

\documentstyle{memsait}
\input epsf.sty
\begin{opening}
\title{ SCIENCE WITH IUE AND THE GROWTH OF AN ASTROPHYSICAL GROUP IN MILAN: 
FROM X-RAY BINARIES TO AGN } 
\author{Renato Falomo$^1$, Laura Maraschi$^2$, Elena Pian$^3$ and Aldo Treves$^4$}
\institute{$^1$Osservatorio Astronomico di Padova, V. Osservatorio 5, 35122 Padova,
Italy\\
$^2$Osservatorio Astronomico di Brera, V. Brera 28, 20121 Milano, Italy\\
$^3$ITESRE-CNR, V. Gobetti 101, 40129 Bologna, Italy\\
$^4$Universit\'a di Como, V. Lucini 3, 22100 Como, Italy\\
}
\date{} 
\end{opening}

\begin{document}

\oddpagefooter{}{}{} 
\evenpagefooter{}{}{} 
\ 
\bigskip

\begin{abstract}

The contribution of an astrophysics  group based in Milan to
the science with the IUE satellite during its almost 20 years
lifetime has focussed on high energy sources, of both galactic
(LMXRB, HMXRB, and black hole candidates) and extragalactic (AGN)
nature.  The results of this long term research and in particular
of the latest multiwavelength campaigns conducted
simultaneously with IUE are reviewed here.

\end{abstract}

\section{Introduction}

IUE was launched in 1978 and has been working for almost two decades.
Its achievements are fully recognized by the astrophysical community.
The impact of IUE observations affected all areas of astronomy from
stars to normal galaxies to Active Galactic Nuclei (AGN) producing
many hundreds of scientific papers and a huge number of Conferences
and Workshops.

Here we want to describe the major scientific achievements obtained by
our group through IUE observations of ``compact objects" in our
galaxy
and beyond, and at the same time we wish to stress the importance of
IUE in the growth and scientific development of Astrophysics in the
Milan area.

In the late sixties and early seventies the astrophysics group in
Milan was led by Beppo Occhialini and Connie Dilworth. The main
interest was in cosmic rays, specifically in cosmic-ray electrons, but
involvement in $\gamma$-ray astronomy had also started.  Optical and IR
astronomy were cultivated by Enrico Tanzi and Oberto Citterio with the
development and use of IR photometers at small telescopes. Massimo
Tarenghi supported Connie Dilworth in conceiving the TIRGO project but
soon moved to Arizona to work with some of the best optical telescopes
available at that time.  Theoretical activity had been started by
Cesare Perola, who was interested in radio sources and AGN, while
Laura Maraschi and Aldo Treves were working on accretion theory with
applications to X-ray binaries and later to AGN.  It was clear that
X-ray emission alone was often insufficient to constrain the physical
conditions of these systems and that, in addition, the high flux
variability suggested the interesting possibility of studying the
correlation of the X-ray with the optical and UV emission.

\subsection{ Why did we start an observational program with IUE? }

The IUE observatory, with its generous opportunities for guest
observers and output calibrated digital spectra that did not
discourage even theoreticians, gave us the possibility of accessing
first class data.  At the same time it favored the coalescence of the
group since the theoreticians needed expertise in classical astronomy
and the "astronomers" were interested in new theoretical motivations
to develop long ranging programs.  Moreover, it provided the
opportunity for us all to collaborate with leading international
groups.  As a result, a strong, long lasting and productive team was
established in the Milan area, which worked with pleasure and
enthusiasm supported by common scientific interests and friendship.
Enrico Tanzi was the leader and the heart of this team. He became
later Deputy Chairman of the IUE Telescope Allocation Committee.

Originally two lines of research were started. One on AGN,
concentrating on 3C 273 and NGC 4151, led in Milan by Cesare Perola,
Enrico Tanzi and Massimo Tarenghi. This program was extremely
successful, becoming an example of effective long term European
collaboration, recall among other participants, J. Clavel, M. Penston,
M.-H. Ulrich and W. Wamsteker.  Later Perola moved to Rome and
Tarenghi to ESO.
 
The second line, focusing on X-ray binaries and cataclysmic variables,
was started by Tanzi and Treves, in collaboration with Pier Luigi
Bernacca at Padova-Asiago.  Also in this case international
collaboration was sought.  In particular, we interacted with the groups
of Bob Wilson at UCL and Andrea Dupree at CFA, who were leading figures
of the IUE project.  Within the collaboration the first UV observations
of the black hole candidate Cyg X-1 and of the magnetic cataclysmic
variable AM Her were performed. They were the object of the thesis of 
Lucio
Chiappetti, who joined permanently the Milan IUE team and became the key
person in managing the local data analysis facilities. 
 
With the second announcement of opportunity of IUE Laura Maraschi joined
the group.  Under the influence of the first results from the X-ray
satellite "Einstein", besides extending the line of research on compact
galactic objects, we began a program on AGN and in particular on BL Lac
Objects, which from interpolation between the optical and X-ray fluxes
were expected to be UV emitters. From the point of view of the accretion
theory, the transition from X-ray binaries to AGN was a rather natural
one. The program, which lasted for the entire duration of IUE, started
with the first UV study of PKS 2155-304, one of the prototypes of X-ray
bright BL Lac objects and one of the brightest AGN in the sky from the
optical to the X-ray band. First results were published in Nature
(Maraschi et al. 1980a), and led to a very fruitful collaboration with
Meg Urry, who had similar interests in the US. 

The IUE data were for us the first step to undertake programs of
multiwavelength observations of our selected targets, aimed at the
study of the broad band spectral flux distribution and variability
thereof, which proved to yield significant constraints on the
emission models. In particular, numerous coordinated observations
were performed with EXOSAT in X-rays and with optical and IR
facilities at ESO. Here the key persons were Enrico Tanzi, and Renato
Falomo. On the systematics of IUE study of BL Lac objects important
contributions came also from  Gabriele Ghisellini and, more recently,
from Elena Pian. 

In the following we will describe some of the  results
obtained with the above programs.  We divide the paper in three main
parts: Section 2 deals with compact galactic sources; Section 3
focuses on the shape of the continuum of BL Lac objects and its
variability from the IR to the UV band; Section 4 reports on the
latest multiwavelength observations involving IUE for two prototypical
blazars: PKS 2155-304 and 3C 279.

We just mention additionally that the first detection of low redshift
intergalactic Ly$\alpha$ absorptions was obtained by coadding a large
number of IUE exposures of PKS 2155-304 (Maraschi et al. 1988).
  
\section{Compact galactic objects}

At the time of launch of IUE, X-ray binaries were one of the main
themes of high energy astrophysics and possibly the main field of
interest of X-ray astronomy. It was recognized that these systems were
powered by accretion of a collapsed object. In most cases this was
supposed to be a neutron star, the presence of which was apparent in
X-ray pulsators. At that time, only one system was assumed  to contain a
black hole, namely Cyg X-1. The discovery of X-ray emission from AM
Her demonstrated the affinity of cataclysmic variables, where
accretion occurs onto a white dwarf, with standard X-ray binaries. The
importance of a distinction of X-ray binaries in low and high mass
systems (LMXRB, HMXRB), depending on the mass of the companion star,
was starting to be recognized.

With this premise, it is no wonder that XRB were among the first
targets of IUE, and some of them were observed within the performance
verification phase of the satellite (Dupree et al. 1978). Immediately
after, a collaborative effort among various groups was established to
optimize the observations with IUE of Cyg X-1, Sco X-1, Vela X-1 and
Her X-1. Later on, the systematic study of XRB and CVs was conducted
by individual groups during the entire lifetime of the satellite.

In Milan, the effort was focussed on both CVs and XRBs. Among CVs
we concentrated on "magnetic CVs" polars, and intermediate polars,
part of them observed simultaneously with EXOSAT (e.g., Tanzi et
al. 1980, Chiappetti et al. 1989, Sambruna et al. 1992). Here we
report briefly only on three XRBs, Cyg X-1 for which we led the
international collaboration, and Cyg X-2 and LMC X-3 for which we
produced the first IUE spectra. 

\subsection{Cyg X-1}

The system is a prototype of HMXRB.  The primary is an O supergiant,
which dominates the optical UV emission.  The O star produces a large
wind, which is supposedly responsible for the mass transfer to the
black hole.  An accretion disk, if present, does not show up in
optical/UV.

The main result of our IUE observations (Treves et al. 1980) was the
discovery of the modulation of resonance lines with the 5.6 days
orbital period. The absorptions were sinusoidally modulated with a
minimum at the inferior conjunction of the X-ray source. The
observation was a clear demonstration of the ionization of the stellar
wind by the X-ray source for wind fed XRB.  Indeed, this modulation
was theoretically predicted by Hatchett and Mc Cray (1977), and was
shown most clearly with IUE in the case of Vela X-1 by Dupree et
al. (1980) and for Cyg X-1 by us.

\subsection{Cyg X-2}

This system is one of best known LMXRB with a period of 9.8d.  The
source is weak for IUE, but its detection was per se a demonstration
of the existence of a UV source much stronger than the non collapsed
star. The UV flux is most probably generated by the re-processing of
X-rays in the accretion disk. In fact, this allowed us to estimate the
geometrical thickness of the accretion disk, which turned out to be
larger than expected, but consistent with that deduced from the
absence of heating of the primary.

The line spectrum of Cyg X-2 is very complex, containing both emission
and absorption lines, clearly indicating a superposition of emitting
regions with different physical conditions.  The prominence of the NV
emission line over CIV is unusual, probably indicating an anomalously
high abundance of nitrogen associated with the evolutionary history of
the
system (Maraschi et al. 1980b, Chiappetti et al. 1983).

\subsection{LMC X-3}

This is one of the three persistent X-ray sources known thus far, with
firm evidence for the presence of a black hole.  It was shown to be a
black hole candidate by Cowley et al. (1983), who discovered the
orbital period of 1.7 d and measured the mass function. Triggered by
this work, we observed the source with EXOSAT and IUE (Treves et al.
1988a,b). 
Also this source is at the limit of IUE detectability.  Still it
was shown to be strongly variable. At the minimum the optical and UV
emission was consistent with that expected from the primary star (BV).
Subtracting the primary from the spectrum at maximum flux we found a
spectral flux distribution, which could be fitted by a black body
(see Fig. 1). This allowed us to establish a first rough inclination of
the system.

\begin{figure}[h]
\epsfysize=6cm 
\epsfxsize=8cm 
\hspace{3cm}\epsfbox{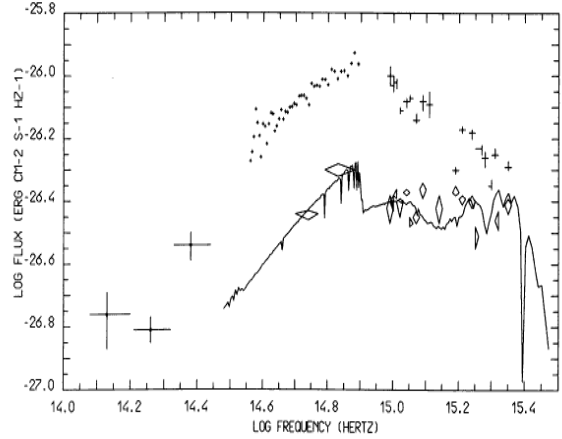} 
\caption[h]{IR to UV flux distribution of LMC X-3. Crosses correspond 
to observations of 1987, diamonds (UV) to March 1986. 
Fluxes in the V and B bands correspond to the lower state of the source.
The continuum tracing represents the Kurucz model of a reddened B3-V 
star.}
\end{figure}

\section{Spectral Continua of AGN observed by IUE}

The use of IUE in the study of AGN has been critical in that the UV
spectral range, unlike the optical and IR, is modestly contaminated by
stellar light of their host galaxies or by thermal emission from dust
surrounding the nucleus.  Moreover, the spectral energy distributions
of AGN often present a local maximum close to the UV wavelengths,
which makes the study of this energy band crucial for understanding
the main energy output.

Our group has predominantly contributed to the investigation of
radio-loud AGN using IUE, particularly blazars, and, to a lesser
extent, of radio-quiet AGN like Seyfert galaxies and quasars.  We
just recall here the early investigations of UV and X-ray
variability of MCG 8-11-11 (Treves et al. 1990), of PG 0026-129
(Treves et al. 1988c) and of 3C 120 (Maraschi et al. 1991) for which 
simultaneous IUE and EXOSAT observations were performed.  In the
case of Kaz 102 (Treves et al. 1995), a systematic IUE campaign was
organized during the ROSAT All Sky Survey which monitored the
source for several months. All these observations indicated the
intricacy of the relation between the emission in the UV and X-ray
band, and the need of very long exposures in the two bands to
clarify such a relation.

Blazars include Highly Polarized and Optically Violently Variable
Quasars (HPQ-OVV) as well as BL Lac objects, characterized by the
absence of strong emission lines, by lower redshift and lower
luminosity (Urry \& Padovani 1995). Their common properties are the
presence of a dominant radio core with flat radio spectrum, ($-0.5 <
\alpha < 0.5$, where $F_\nu \propto\nu^{-\alpha}$), strong radio
and/or optical polarization ($>$ 3\%), and rapid variability at all
frequencies.  The radio to optical continuum is thought to be produced
by synchrotron radiation of high energy particles in a relativistic
jet, pointing at small angle to the line of sight. The observed
radiation is then strongly "amplified" by relativistic beaming. Indeed,
some blazars present the highest apparent bolometric luminosities
among all AGN.

\subsection {Broad Band Spectra and Variability of Individual Sources}

To precisely characterize the multiwavelength spectral shape of
blazars, simultaneous observations are needed at the different
wavelengths (i.e., less separated in time than the typical variability
time scale at each wavelength).  The flexibility of the IUE scheduling
has allowed us to organize numerous observing campaigns of blazars at
UV and other wavelengths (notably optical and X-ray) at single epochs,
to measure and compare the fluxes and spectral slopes in different
spectral ranges, and therefore reconstruct the overall energy
distribution.

Examples of multiwavelength spectra obtained during
different brightness
states are given for PKS~0537--441 (Tanzi et al. 1986) and PKS~2155--304
(Treves et al. 1989). The former source ($z = 0.894$)  exhibited a
brightening by a factor $\sim$ 2 in the whole range from near-IR tu UV
whitout any significant change of slope (Fig. 2).  Similar behaviour was
seen in
the BL Lac object PKS~2155--304 ($z = 0.116$), for which the spectral
flux distribution was extended up to X-ray frequencies through EXOSAT
observations (Fig. 3).

\begin{figure}
\epsfysize=6cm 
\epsfxsize=8cm 
\hspace{3cm}\epsfbox{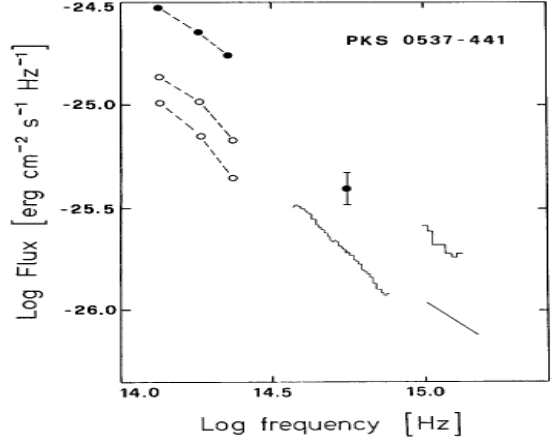} 
\caption[h]{
Spectral flux distribution of PKS 0537--441 in two different
brightness
states.  The filled circles and the steplike line in the UV range
represent measurements obtained during the brightening of 
February 1995. The other data represent a lower brightness state.  Open
circles in the near-IR are from Allen et al. (1982).  The steplike
line
in the optical represents our observations of November 1984 and the
continuum line in the UV is from the 1980 observations by Maraschi et
al. (1985).}
\end{figure}

\begin{figure}
\epsfysize=6cm 
\epsfxsize=8cm 
\hspace{3cm}\epsfbox{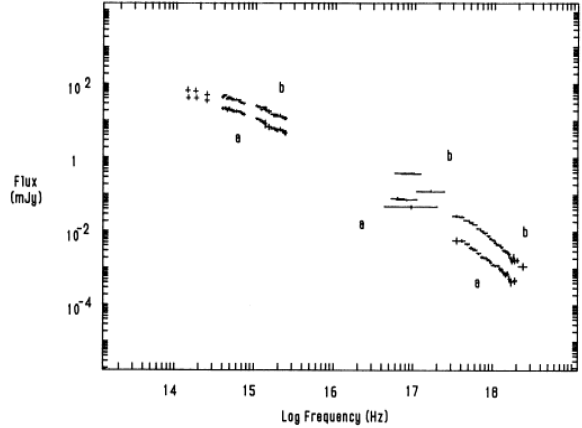} 
\caption[h]{
Spectral flux distribution of PKS 2155-304 for October 1983 ({\it
lower state} labelled a)  and 1984 November 11 ({\it higher state}
labeled b).  The spectral shape increasingly
steepens towards higher frequencies, but its slope does
not change significantly between the two luminosity levels.}
\end{figure}

The simultaneous UV, optical and near-IR spectral flux distributions
of 11 blazars selected for being the most strongly polarized among
those of the complete samples of Impey \& Tapia (1988), when properly
corrected for Galactic reddening and decomposed from the stellar host
galaxy contribution, are well described by a single power-law (Pian
et al. 1994). This suggests that the radiative process which is
responsible for the emission in the frequency range 8$\times 10^{13}$
- 2$\times 10^{15}$ Hz (synchrotron radiation) takes place inside a
single emitting volume.

Among these sources, PKS~0521--365, PKS~2005--489 and PKS~2155--304
were detected in $\gamma$-rays by the EGRET instrument onboard the CGRO
satellite. For PKS~0521--365, simultaneous UV and X-ray (EXOSAT)
observations in November 1983, combined with radio, IR, optical and
$\gamma$-ray data taken at different epochs, show that the spectral flux
distribution declines from the IR to the UV range (energy index
$\alpha_{opt-UV} = 1.4 \pm 0.1$), but {\it hardens} at higher
frequencies (Pian et al. 1996).


The remarkably flat X-ray spectrum ($\alpha_X = 0.68 \pm 0.12$
in the range 0.2--7 keV) and the detection of the source in 
$\gamma$-rays suggest a significant contribution by a mechanism other 
than synchrotron. Given the presence of high energy electrons and 
the high radiation density in these sources it is natural to invoke 
the inverse Compton mechanism to account for the hard X-ray emission.

The spectral energy distribution can be interpreted with a
relativistic jet model in which the near-IR-to-UV radiation is
produced via the synchrotron process in the internal part of the jet
(close to the nucleus) and the radio emission comes from the external
part (Ghisellini et al. 1985).  Various arguments indicate that the
jet in this source is not closely aligned with the line of sight
($\theta \sim 30^{\circ}$), implying that the observed radiation is
only weakly boosted.  A direct estimate of the importance of different
sources of photons available for upscattering suggests that in this
source the synchrotron photons are likely to be the seeds for the
inverse Compton process.

PKS~0521--365 also exhibits a strong Ly$\alpha$
emission line (Scarpa, Falomo, \& Pian 1995) which does not vary
significantly, in contrast with the remarkable variations of the UV
continuum.

 
\subsection {Comparative Studies }

A comparative study of the spectral shapes of BL Lac objects in the UV
yielded the first evidence of a systematic difference with respect to
quasars and of a correlation of the UV spectral index with the broad
band radio to X-ray index (Ghisellini et al. 1986). This result
supported the non thermal nature of the UV emission from BL Lac
objects, in contrast with the thermal origin of the UV bump in quasars
(Bregman, Maraschi, \& Urry 1987) .

The question was later reconsidered using the IUE archival spectra of
47 blazars and of a substantial fraction of {\it Palomar Green} bright
quasars (Pian \& Treves 1993).

The average spectral index in the range 1200--3000 {\rm \AA} was found
to be $\alpha_{UV} \simeq$ 1, with HPQ being steeper (($\alpha_{UV}
\simeq$ 1.4) than radio-strong BL Lacs (defined by $\alpha_{radio-x} >
0.8$), and these in turn being steeper ($\alpha_{UV} \simeq$ 1.2) than
radio-weak BL Lacs ($\alpha_{UV} \simeq$ 0.7).  Such distinction,
which is also seen in the total luminosity, was originally proposed to
be related to the jet viewing angle (Ghisellini \& Maraschi
1989). More recently, a better knowledge of the overall radio to X-ray
continuum led to suggest intrinsic physical differences (Sambruna,
Maraschi, \& Urry 1996).

The UV spectrum of HPQ and radio-strong BL Lacs differs significantly
from that of PG quasars, for which the average spectral index is
$\alpha_{UV} \simeq$ 0.8.  The flatter UV spectrum of bright nearby
quasars is probably due to the contribution of the thermal component
(blue bump), which is less prominent (or absent) in blazars.

The UV variability of these blazars was studied by Treves \& Girardi
(1990), who found a correlation of UV variability with UV luminosity
and a significantly larger variability at shorter UV wavelengths
($\sim$1500\AA) than at longer ones ($\sim$2500 \AA), in agreement 
with the
overall variability-energy correlation seen in broad-band 
blazar spectra (Ulrich, Maraschi, \& Urry 1997). 


\section{Blazar Multiwavelength Campaigns}

It was clear after the results above that at least some blazars vary
extremely rapidly in the UV and that quasi-simultaneous snapshots of
the UV to X-ray energy distribution were insufficient to probe the
correlation between the variability in two wavelength
ranges. Simultaneous light curves in X-rays and UV and possibly other
wavelengths were and are still needed to address physical models of
variability. Despite the long lifetime of IUE well sampled
multiwavelength data were obtained in a limited number of cases (see
for reviews Wagner \& Witzel 1995; Ulrich et al. 1997).  Here we
will
discuss two sources for which many data have been obtained which
helped us to make progress in the understanding of the blazar
phenomenon. The first is the BL Lac object PKS 2155-304, one of the
brightest blazars in the UV and soft X-ray sky. The second is the
superluminal quasar 3C 279, the first and one of the brightest blazars
detected in $\gamma$-rays.

The discovery by the EGRET instrument on board CGRO that blazars emit
copious fluxes of $\gamma$-rays changed substantially our view of the
energy distribution. These observations revealed a new spectral
component, a hint of which was given by the flat X-ray spectra of some
blazars, with power comparable and sometimes dominant with respect to
the previously known synchrotron component.

The spectral energy distributions of blazars from the radio to the
$\gamma$-ray range can now be described in a general way as composed of
two broad peaks in the power per decade, corresponding to the
synchrotron and inverse Compton emission, respectively.  Blazars of
different luminosities appear to lie along a spectral sequence
(Fossati et al. 1998) whereby both peaks fall at higher frequencies
for lower luminosity objects.  Thus highly luminous objects are "red",
meaning that the first peak falls at frequencies smaller than the
optical range and therefore the optical-UV continuum is steep, while
low luminosity objects are "blue" having peak frequency beyond the UV
range and therefore flat (blue) optical-UV continua.  For red
blazars the second spectral component peaks within or below the EGRET
(0.1-10 GeV) range, while for blue blazars it peaks beyond the EGRET
range.  PKS 2155-304 and 3C 279 are prototypes of blue and red
blazars, respectively.
 
From IR frequencies upwards the synchrotron emission should be
optically thin and could be produced in a single zone of the jet,
allowing the adoption of a homogeneous model.  The second spectral
component
(peaking in $\gamma$-rays) is thought to be produced by high energy
electrons upscattering soft photons via the inverse Compton
process. The seed photons for upscattering could be the synchrotron
photons themselves (synchrotron self Compton model, SSC) or photons
outside the jet (external Compton model, EC) possibly produced in an
accretion disk or torus and scattered or reprocessed by the
surrounding gas (e.g., Sikora 1994; Ulrich et al. 1997, and
references therein).

If the same region is responsible for the two spectral components,
irrespective of the nature of the seed photons, {\it emission at the
two peaks must derive from the same high energy electrons}.  Therefore
a change in the density and/or spectrum of those electrons {\it is
expected to cause correlated variability at frequencies close to the
two peaks.}  Measuring the two peaks simultaneously is thus an
essential step for determining the physical parameters of the emission
region and studying the variability of the spectra around the peaks
yields unique insight into the mechanisms of particle acceleration and
energy loss in the jet.  The variability correlation should enable
disentangling the contribution of different sources of seed photons
(SSC vs. EC).


\subsection {PKS 2155--304}

PKS 2155-304 is one of the brightest sources in the extragalactic sky
at UV and soft X-ray wavelengths. As such, it was repeatedly observed
in many wavebands (Maraschi et al. 1986, Urry et al. 1988, Treves et
al. 1989, Edelson et al. 1991).  Two major intensive multiwavelength
campaigns were organized on this source.  The first was based on 5
days of quasi-continuous coverage with IUE and 3.5 days with ROSAT
(Urry et al. 1993, Brinkmann et al. 1994, Edelson et al. 1995).  The
UV and X-ray light curves showed rapid variations of moderate
amplitude ($\simeq 20\%$), strongly correlated in the overlapping time
intervals, with the soft X-rays leading the UV variations by no more
than 2-3 hours.  Therefore variability was essentially "achromatic",
that is independent of wavelength, ruling out an accretion disk as the
origin of the UV continuum.  Over the whole monitoring period of about
1 month the UV intensity increased by a factor 2, the same as observed
in the optical and IR.


\begin{figure}
\epsfysize=6cm 
\epsfxsize=8cm 
\hspace{3cm}\epsfbox{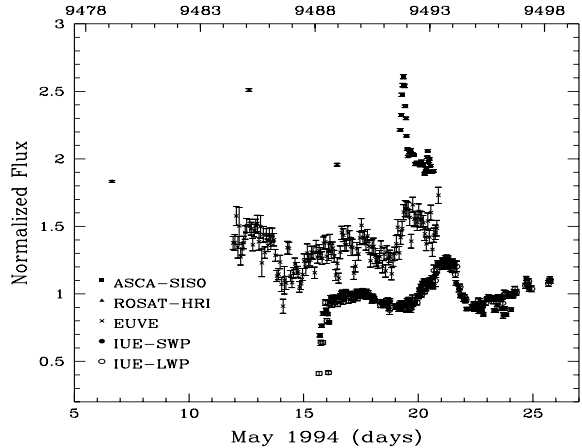} 
\caption[h]{Normalized X-ray, EUV and UV light curves of PKS 2155-304
from the second intensive multiwavelength monitoring campaign, in May 
1994.
The ASCA data show a strong flare, echoed one day later in the EUVE 
and two
days later in the IUE light curves. The amplitude of the flares 
decreases and
the duration increases with increasing wavelength (from Urry et al. 
1997).}
\end{figure}

A second, longer campaign, with $\simeq$ 10 days of IUE, $\simeq$ 9
days of EUVE, 2 days of ASCA, and three short ROSAT observations, took
place in May 1994 (Pesce et al. 1997, Pian et al. 1997, Urry et
al. 1997) and caught a well defined flare (factor 2.5) during the two
days of ASCA observations, which can be most plausibly related with
the smaller amplitude intensity peaks (35~\%) seen at later times in
the EUVE and IUE light curves.  The light curves from this campaign
are shown in Figure 4.  The X-ray flare appears to lead the EUVE and
UV events by 1 and 2 days, respectively, an order of magnitude longer
than the lag detected in the previous campaign. Within the ASCA data
the 0.5-1 keV photons lagged the 2.2-10 keV photons by 1.5
hours.

Note how the UV light curve is well defined and how the differences
between the LWP and SWP normalized fluxes are small, except for the
initial extremely fast and deep minimum. This event is discussed in
detail in Pian et al. (1997) and is not completely understood.

Despite the differences, these two campaigns give the first evidence
that the variations from 5 eV to 10 keV are correlated on short time
scales and that the emission at higher frequencies leads the one at
lower frequencies. The flare event
observed during the second campaign has characteristics as expected
from the propagation of a shock wave along an inhomogeneous
relativistic jet.  The cause of the flare in this model is a
propagating disturbance affecting different regions at different
times.

\subsection{3C 279}

3C~279 ($z = 0.54$)  was the first blazar discovered 
to emit strong and variable $\gamma$-rays (Hartman et al. 1992).

There have been several multi-wavelength observations of this blazar
(Maraschi et al. 1994, Hartman et al. 1996, Wehrle et al. 1998).  In
particular, 3C~279 was monitored with IUE and ROSAT for three weeks
between December 1992 and January 1993, simultaneously with
$\gamma$-ray observations by EGRET, and with coordinated optical
observations. At that epoch the intensity of 3C~279 was at a
historical minimum at all measured wavelengths above the sub-mm ones.
The variability amplitude of the continuum with respect to the bright
state of June
1991 increased with frequency from the IR to the UV. It is
possible that the residual UV emission contains a substantial
contribution from an accretion disk (see Pian et al. 1998).
This thermal radiation, which is
expected to vary with more modest amplitude than the  
beamed synchrotron UV
emission, might be responsible for powering the strong, not
significantly variable, Ly$\alpha$ emission line, visible in all
archival IUE SWP spectra of 3C~279 (Koratkar et al. 1998). 

Regarding the second spectral component the variability amplitude was
small in the soft (ROSAT) X-ray range but extremely large in
$\gamma$-rays, larger than in any other waveband. 
This showed that the inverse Compton emission had varied 
much more than the synchrotron
one, an effect expected in the SSC model. 
A similar behaviour could be
reproduced in the EC model only if the ambient photons 
had varied in a correlated fashion with the electrons in the jet.

\begin{figure}[h]
\epsfysize=10cm 
\epsfxsize=8cm 
\hspace{3cm}\epsfbox{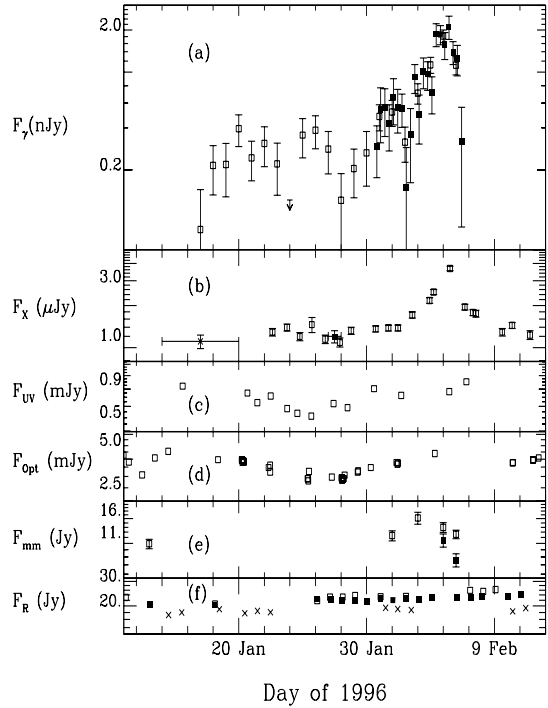} 
\caption[h]{Multiwavelength light curves of 3C~279 during the EGRET 
campaign 1996
January 16-February 6 (from Wehrle et al. 1998): {\it (a)} EGRET fluxes
at $>$100 MeV binned
within 1 day (open squares) and 8 hours (filled squares) (referred to 400
MeV); {\it (b)} X-ray fluxes at 2 keV:
besides the RXTE data (open squares), the isolated ASCA (filled square)
and ROSAT-HRI (cross) points are reported with horizontal bars indicating
the total duration of the observation; {\it (c)} IUE-LWP fluxes at 2600
\AA; {\it (d)} ground-based optical data from various ground-based
telescopes in the R band;  {\it (e)} JCMT photometry at 0.8 mm (open
squares) and 0.45 mm (filled squares); {\it (f)} radio data from
Mets\"ahovi at 37 GHz (open squares)  and 22 GHz (filled squares), and
from UMRAO at 14.5 GHz (crosses).  Errors, representing 1-$\sigma$
uncertainties, have been reported only when they are bigger than the
symbol size.}
\end{figure}

A second intensive multiwavelength monitoring, based on a 3-week
coordinated program involving CGRO, XTE and IUE observations, took
place in January-February 1996 (Wehrle et al. 1998). 
A huge flare was observed
in $\gamma$-rays and a similar but less extreme event was seen by
XTE. The X and $\gamma$-ray peaks were simultaneous within the one day
uncertainty.  The IUE light curve at 2600 \AA\ (Fig.~5) is reasonably
well sampled during the first part of the campaign but not toward the
end, when the flare occurred and observations of 3C~279 were extended
as Target of Opportunity.  It shows a broad minimum at $\sim$ 25-26 
January, followed by a rise of almost a factor 2, but with a three day
gap before and up to the $\gamma$-ray peak. Due to the faintness of
the source and to the scattered light problem during the final stages
of the IUE mission, systematic errors may affect some of the IUE
observations.  Despite the non optimal sampling, the IUE observations
provide important constraints, as discussed below.

In Figure 6 the broad band energy distributions obtained at the 
flare peak and in a preflare state (1996 January 25-26) are compared 
with those of the discovery epoch (June 1991) and with the low 
state of January 1993. 

Comparing the flare and preflare states it is clear that the high
energy SED (X-ray to $\gamma$-ray) is harder at the flare peak, as
implied by the larger amplitude of the $\gamma$-ray variation.  From
IR to UV frequencies the flare vs preflare variations are much smaller
than in X-rays and $\gamma$-rays.  This has important consequences for
theoretical models, as the relative variability in the synchrotron and
inverse Compton components can indicate the origin of the seed photons
that may be upscattered to the $\gamma$-ray band.

\begin{figure}
\epsfysize=6cm 
\epsfxsize=8cm 
\hspace{3cm}\epsfbox{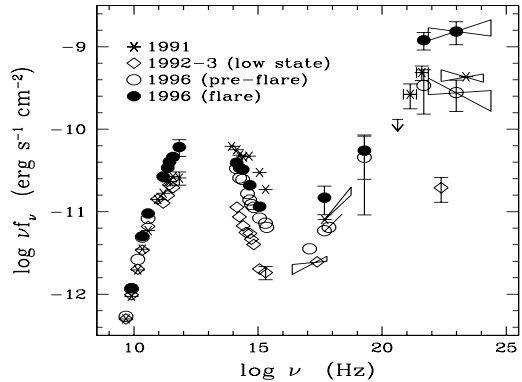} 
\caption[h]{Radio-to-$\gamma$-ray energy distributions of 3C~279 
in different states:
stars refer to the high state of June 1991 (Hartman et al.  1996);
diamonds to the low state of December 1992-January 1993 (Maraschi et al.
1994);  open and filled circles represent the pre-flare and flaring state
in January-February 1996, respectively (Wehrle et al.  1998).  The
near-IR, optical, UV and soft X-ray data have been corrected for Galactic
extinction.  Errors have been reported only when they are bigger than the
symbol size.} 
\end{figure}

The amplitude of the $\gamma$-ray variation during the 1996 outburst
(factor 10) is more than the square of the observed IR-optical flux
change (factor $\sim 1.5$) which is a severe difficulty for both the
SSC and EC scenarios. Possible ways out are the following: (i)
different emission zones could contribute to the IR-optical flux,
diluting the intrinsic variation associated with the $\gamma$-ray
flaring region; (ii) the synchrotron peak corresponding to the
$\gamma$-ray peak may fall at frequencies lower than IR, where
adequate variations could have occurred; (iii) an alternative
possibility is the "relativistic mirror" model of Ghisellini and Madau
(1996), combining advantages of both the SSC and EC models.  While we
cannot completely rule out the SSC scenario or, for similar arguments
the "standard" EC scenario, the mirror model requires very special
conditions and is under discussion (Boettcher \& Dermer 1998).

\section{ Conclusions}

The years spent motivating, obtaining, analysing and discussing IUE 
data were a vital phase in the evolution of astrophysical research in 
the Milan area and were certainly scientifically very productive. 

Because of our interest in collapsed objects, we exploited among
the
capabilities of IUE those that were better matched to the joint
study of the UV and X-ray emission. The low resolution mode was
used since most of the objects were weak but also because it
allowed a good measure of the spectral shape of the continuum,
which was in many cases an essential objective of the
investigation. 

In retrospect, we can say that our contributions to the study of the
interrelations between the UV and the X-ray emission were very
successful. In galactic systems they helped localizing different
emission regions and reconstructing the matter flow towards the
central object. In Seyfert galaxies and quasars the problems that
could be addressed were similar although on different scales. It is
worthwhile mentioning that the issue of the dominance of the UV
{\it bump} or of the hard X-ray emission in the broad band energy
distribution of AGN is still not completely assessed for different
classes of sources. 

For blazars, the continuity or ``discontinuity" of the UV and X-ray
spectral shapes were the first evidence of the presence of two
distinct emission components later spectacularly confirmed by the
$\gamma$-ray observations.  The recent multiwavelength campaigns
have shown conclusively that both components derive from the same
high energy electrons accelerated in a relativistic jet. 
Understanding the nature of the seed photons upscattered to
$\gamma$-ray energies and probing the nature of the acceleration
mechanisms seem reachable goals for further multiwavelength
observations. 

In our experience a lasting lesson from IUE has been its scheduling
flexibility, which allowed for the first time the study of
correlated optical-UV-X-ray variability in a large number of
astrophysical situations, some of which illustrated above.

The scientific importance of such correlation studies has been
widely recognized, and indeed new X-ray missions are designed with
the addition of small optical-UV telescopes (XMM, INTEGRAL,
SPECTRUM - X $\gamma$).  However these ``complementary" instruments
have little spectroscopic capabilities and limited wavelength
ranges even for studies of the continuum.  At the other extreme, 
the
UV capabilities of HST are unrivalled but its rigid scheduling
poses
great difficulties for this kind of observations. It seems
therefore that the IUE heritage still needs to be fully recognized
and pursued by future missions.




\end{document}